%% file: SF-CKM-procN.tex
\documentclass[12pt]{article}
\usepackage{epsfig}
\usepackage{amsmath}

\input econfmacros.tex

\def\Title#1{\begin{center} {\Large {\bf #1} } \end{center}}

\begin{document}

\Title{Theory of $B\to \tau \nu $ and $B\to D^{(*)}\tau \nu $ }

\bigskip\bigskip

\begin{raggedright}  
{\it  Svjetlana Fajfer\index{Fajfer,S.} and Ivan Ni\v sand\v zi\' c \\
Institut Jo\v zef Stefan, \\
Jamova 39, P. O. Box 3000,
1001 Ljubljana, Slovenia, \\
and \\
Department of Physics, University of Ljubljana, \\
Jadranska 19, 1000 Ljubljana, Slovenia}
\bigskip\bigskip
\end{raggedright}

\section{Introduction}
B physics plays an important role in testing the Standard Model (SM) at low energies.  Recent experimental results on CP violation in the $B_s\to J/\psi \phi$ decay and the decay rate for $\mathcal B(B_s \to \mu^+ \mu^-)$ severely constrain possible New Physics (NP) contributions to these observables. 
Analogously, measurements of $ \mathcal B(B\to \tau \nu)$ and  $\mathcal B(B\to D^{(*)} \tau \nu)$  probe the possible impact of beyond SM  physics in the leptonic and semileptonic $B$ decays.
Within the SM, these decay modes are important since they contribute in obtaining precise values of the $|V_{ub}|$ and $|V_{cb}|$ together with the relevant hadronic decay constants or  form factors. 
For example due to the large mass of the $\tau$, semileptonic decays $B\to D^{(*)} \tau \nu$ are sensitive to additional form factors,  which are unimportant in the corresponding $B$ decays with light leptons in the final state.  In addition these tauonic decay modes represent sensitive tests of  lepton flavor universality (LFU) in charged current interactions. 
The most recent world average of the leptonic $B\to \tau \nu$ branching fraction measurements
$\mathcal B(B^-\to \tau^- \bar \nu_\tau) = (11.4\pm2.3)\times 10^{-5}$~ \cite{1}
deviates somewhat from its SM prediction (with the $V_{ub}$ CKM element taken from the global fit~\cite{Charles}). In contrast, the measured exclusive semileptonic $b\to u\ell\nu$ transition branching fraction
$\mathcal B(\bar B^0\to \pi^+ \ell^- \bar \nu_\ell) = (14.6\pm0.7)\times 10^{-5}$~\cite{BaBarpi,HFAG} 
is consistent with SM result used by UTFit, $\mathcal B(\bar B^0\to \pi^+ \ell^- \bar \nu_\ell)=(1.31\pm0.06\pm 0.06)\times 10^{-4}$~\cite{Charles}. 
Furthermore, the BaBar collaboration has recently published results of their  measurements of $B\to D^{(*)}\tau \nu$ branching fractions normalized to the corresponding $B\to D^{(*)}\ell \nu$ modes (with $\ell=e,\mu$)~\cite{BaBarBD} 
\begin{subequations}
\begin{align}
\mathcal R^*_{\tau/\ell}& = {\mathcal B(B\to D^* \tau^- \bar \nu_\tau)}/{\mathcal B (B\to D^{*}\ell^- \bar \nu_\ell)} = 0.332 \pm 0.03 \,,
 \label{eq:Rstar}\\
\mathcal R_{\tau/\ell} & = {\mathcal B(B\to D \tau^- \bar \nu_\tau)}/{\mathcal B(B\to D^{}\ell^- \bar \nu_\ell)} = 0.440\pm 0.072\,,  \label{eq:R}
\end{align}
\end{subequations}
where the statistical and systematic uncertainties have been combined in quadrature.
Both values $\mathcal R^*_{\tau/\ell} $ and $\mathcal R_{\tau/\ell}$  are in agreement  with previous measurements, 
 but larger  (at $3.4\sigma$ significance when combined) than the SM values $\mathcal R_{\tau/\ell} ^{*,\rm SM} = 0.252(3)$ and $\mathcal R_{\tau/\ell} ^{\rm SM} = 0.296(16)$ \cite{FKN}.  If confirmed, these results might point to NP in (semi)tauonic $b \to u(c)$ transitions.

\section{$\bf B\to \tau \nu$}

During the last three years there has been a systematic disagreement between the experimental and SM predicted theoretical values for the branching ratio of $B\to \tau \nu$.
The latest Belle collaboration result  $\mathcal B(B^-\to \tau^- \bar \nu_\tau)= (0.72^{+0.27}_{-0.25} \pm 0.11)\times 10^{-4} $ \cite{Belle1} ameliorates somewhat the enduring tension with the measured value of $\sin 2 \beta$ in the  global CKM fit.  However, the current world average experimental value still deviates from the SM prediction by $2.6 \sigma$  significance if Gaussian errors are assumed~\cite{CKMF}. 

The $B$ meson coupling constant is the only hadronic parameter entering the theoretical branching ratio prediction. The errors of the most recent lattice QCD results are at the level of $5\%$~\cite{Laiho} and already subleading compared to the dominant parametric uncertainty due to $|V_{ub}|$. 
 One can eliminate the $V_{ub}$ dependence completely by introducing  the LFU probing ratio $\mathcal R^{\pi}_{\tau/\ell} \equiv [\tau(B^0)/\tau(B^-)][\mathcal B(B^-\to \tau^-\bar\nu_\tau)/\mathcal B(\bar B^0 \to \pi^+\ell^-\bar \nu_\ell)] = 0.73\pm0.1$.
This is to be compared to the SM prediction of $\mathcal R^{\pi, \rm SM}_{\tau/\ell} = 0.31(6)$~\cite{FKNZ}. 
The measured value is more than a factor of 2 bigger -- a  discrepancy with $2.6\sigma$ significance if Gaussian errors are assumed.


\section{New Physics in $\bf b\to c(u) \tau  \nu$ }
%

The $\tau$  lepton in the final state of the (semi)leptonic $B$ meson decays is particularly interesting due to the large $\tau$ mass which 
allows to probe parts of amplitudes in $B$ meson (semi)leptonic decays  which are not accessible if the final state contains only light leptons.  
Possible  NP effects in the  ratios $\mathcal R^{(*)}_{\tau/\ell}$ and  $\mathcal R^{\pi}_{\tau/\ell}$ can be approached by using the effective Lagrangian approach \cite{FKNZ,FKN}.  The SM Lagrangian is extended with a set of higher dimensional operators, $\mathcal Q_i$, that are generated at a NP scale $\Lambda$ above the electroweak symmetry breaking scale $v= (\sqrt{2}/4G_F)^{1/2}\simeq 174$~GeV
\begin{equation}
\mathcal L = \mathcal L_{\rm SM} +  \sum_a \frac{z_a}{\Lambda^{d_a-4}} \mathcal Q_i + \rm h.c.\,,
\label{eq:Lagr}
\end{equation}
where $d_a$  stand for  the  dimensions of the operators $\mathcal Q_a$, while $z_a$ are the dimensionless Wilson coefficients (below we also use $c_a\equiv z_a (v/\Lambda)^{d_a-4}$). Two restrictions are enforced:  (i) dangerous down-type flavor changing neutral currents (FCNCs) and (ii)  LFU violations in the pion and kaon sectors are not to be generated at the tree level.
 The lowest dimensional operators that can modify 
$R^{(*)}_{\tau/\ell}$ and $\mathcal R^{\pi}_{\tau/\ell}$  are then
\begin{subequations}
\begin{align}
\mathcal Q_{L} &= (\bar q_3 \gamma_\mu \tau^a q_3) \mathcal J^\mu_{3,a}\,, & \mathcal Q^{i}_{R} &=  (\bar u_{R,i} \gamma_\mu  b_R) (H^\dagger \tau^a \tilde H)  \mathcal J^\mu_{3,a}\,,\\ \mathcal Q_{LR} &=  i \partial_\mu (\bar q_3 \tau^a H  b_R) {\textstyle \sum_j} \mathcal J^\mu_{j,a}\,, & \mathcal Q^{i}_{RL} &=  i \partial_\mu (\bar u_{R,i} \tilde H^\dagger \tau^a q_3)  {\textstyle \sum_j} \mathcal J^\mu_{j,a}\,, \label{eq:QRL}
\end{align}
\end{subequations}
where $\tau_a = \sigma_a/2$,   $\mathcal J^\mu_{j,a} =  (\bar l_j \gamma^\mu \tau_a l_j)$, $\tilde H \equiv i \sigma_2 H^*$ and $i,j$ are generational indices. 
We work in the down quark mass basis, where $q_i = (V^{ji*}_{CKM} u_{L,j},d_{L,i})^T$, and charged lepton mass basis, $l_i = (V^{ji*}_{PMNS}\nu_{L,j},e_{L,i})^T$. The requirement that there are no down-type tree-level FCNCs imposes  flavor alignment in the down sector for the operators $\mathcal Q_L, \mathcal Q_{LR}$ and $\mathcal Q_{RL}^i$. 
An additional possibility is to assume~\cite{FKNZ} the presence of  new light invisible fermions, imitating   the missing energy signature of SM neutrinos in the $b\to u_i \tau \nu$ decays. 

\begin{figure}[t]
\centering{
\includegraphics[height=0.35\textwidth]{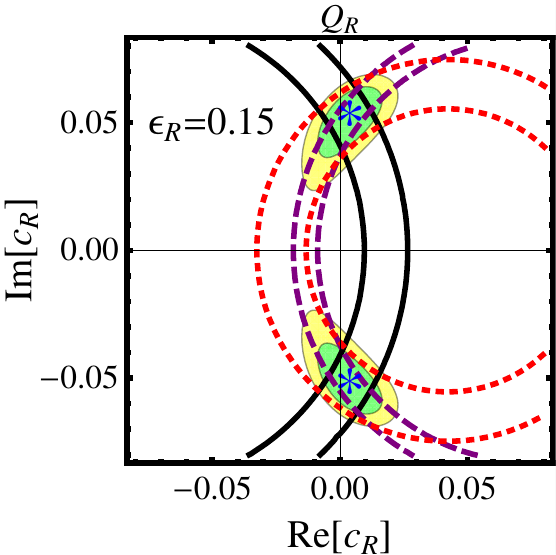}
\includegraphics[height=0.35\textwidth]{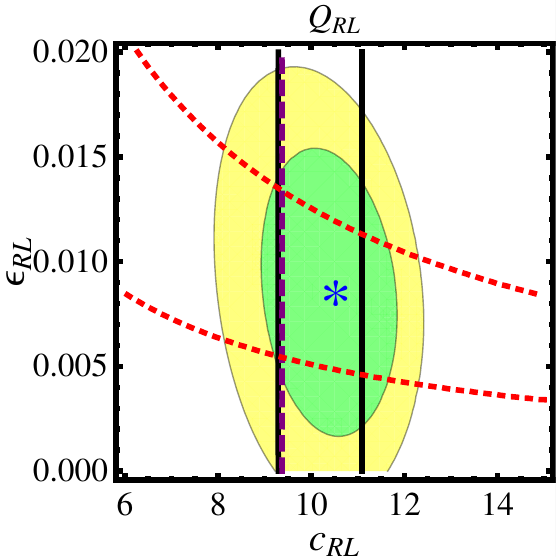}
}
\caption{\footnotesize  Preferred parameter regions for effective operators $\mathcal Q^i_{R}$ (left plot, as a funciton of complex $ c_R$ Wilson coefficient, and $ \epsilon_R$ fixed to the best fit value),  and for $\mathcal Q^i_{RL}$ (right plot, as a function of real $ c_{RL}$ Wilson coefficient and the mixing ratio $ \epsilon_{RL}$).  The best fit points are marked with an asterisk.}
\label{fig:2}
\end{figure}

In the presence of general flavor violating  NP, contributions to $b\to u$ transitions are not generally related to  $b\to c$ transitions. 
In the case of $\mathcal Q^i_{R}$ for example, the SM expectations are rescaled by  $|1- c_{R}/2V_{cb}|^2$ in the case of $\mathcal R_{\tau/\ell}$ and by $|1+ \epsilon_R c_{R}/2V_{ub}|^2$ for $\mathcal R^{\pi}_{\tau/\ell}$. 
The parameters $( c_i, \epsilon_i)$    can be obtained by  fitting  the data  using CKM inputs from the global fit as given in ~\cite{FKNZ}. The results are presented in Fig.~\ref{fig:2}.

Among existing NP models  the two-Higgs doublet models (2HDMs) are obvious candidates to induce the $\mathcal Q_{RL}^i$ operators. 
Unfortunately,  none of the 2HDMs with natural flavor conservation can simultaneously account for the three considered LFU ratios, while in ref. ~\cite{FKNZ} a 2HDM with more general flavor structure has been considered  explaining all the observed deviations.

\section{New Observables in $B \to D \tau \nu$  and $B \to D^* \tau \nu$ }

Recently, the authors of~\cite{Damir}  noticed that the two form factors appearing in $B \to D \tau \nu_\tau$ can be related as  $F_0(q^2)/F_+(q^2) = 1- \alpha q^2$ where the current lattice knowledge of both  form factors allows for a very precise determination of $\alpha=0.020(1)$ GeV$^{-2}$. Based on the regions of $q^2$ directly accessible to the lattice computations, they suggested to measure 
$R^{D}_{\tau,\mu}= \frac{ \mathcal B( B\to D \tau \nu_\tau) }{ \mathcal B( B\to D \mu \nu_\mu)}|_{q^2\leq 8 \rm{GeV}^2} $,
which is precisely predicted in the SM with $R^{D,\rm SM}_{\tau,\mu}= 0.20 \pm0.02$.

In the  decay amplitude of $B \to D^* \tau \nu_\tau$  it is convenient to introduce  helicity amplitudes (for details see ~\cite{FKN}). 
The presence of the pseudo-scalar NP operator only affects the $H_{0t}$ helicity amplitude and can be included as
 \begin{equation}
H_{0t} = H_{0t}^{\rm SM} \left[1 + (g_{SR}-g_{SL}) \frac{q^2}{m_b+m_c}\right]\,.
\label{eq:HNP}
\end{equation}

 The experimental branching fraction measurements of $B\to D^{(*)}\tau\bar\nu_\tau$ decays are systematically above SM predictions. It was found in \cite{FKN} that a fit to the experimental results prefers a non-SM solution with $g_{SL} \simeq -0.9\,$GeV${^{-1}}$. It is important to point out that these NP operators only contribute to the longitudinally polarized $D^*$ ($D^*_L$) in the final state. 



The helicity amplitudes $H_{00}$ and  $H_{0t}$ contribute $D^*_L$'s, leading to a prediction for the longitudinal rate. One can also study the singly differential longitudinal rate ratio $R^*_L(q^2)$ defined analogously to $R^*(q^2)$  as described in~\cite{FKN}. 
 A  simple angular (opening angle) asymmetry is defined as the difference between partial rates where the angle $\theta$ between the $D^*$ and $\tau$ three-momenta in the $\tau-\bar \nu_\tau$ rest-frame  (see.g. \cite{FKN})
is bigger or smaller than $\pi/2$. 
In the decay modes with light leptons, this asymmetry ($A^\ell_\theta$) can be used to probe for the presence of right-handed $b\to c$ currents, since these contribute with opposite sign to $H_{\pm\pm}$ relative to the SM. In the tau modes, it is sensitive only to the real part of NP $g_{SR}-g_{SL}$ contributions and thus provides complementary information compared to the total rate (or $R^*$). 
On the other hand, the inclusive asymmetry $A_\theta$ integrated over $q^2$ is very small in the SM with $A_{\theta,\rm SM} = -6.0(8)\%$; for our NP benchmark point we obtain $A_{\theta,\rm NP} = 3.4\%$\,, but even values as low as $-30\%$ are still allowed. In~\cite{FKN}  it was found that the tau spin asymmetry, defined as $A_{\lambda}(q^2)=[d\Gamma_\tau/{dq^2}(\lambda_\tau=-1/2)-d\Gamma_\tau/{dq^2}(\lambda_\tau=1/2)]/ [{d\Gamma_\tau}/{dq^2}]$, where $\lambda_\tau=\pm 1/2$ are tau helicities defined in $\tau\nu_\tau$ center of mass frame, can provide additional useful information.
%
\section{Summary}
%
Within an effective field theory approach  the most general lowest dimensional contributions to helicity suppressed (semi)leptonic $b \to c(u)$ transitions have been considered and found that a precise study of the exclusive decay mode $B\to D^* \tau \bar \nu_\tau$ could clarify the possible existence of such non-SM physics.  Recent experimental value   $\mathcal B(B \to \tau \nu) $  is closer to SM prediction and more data on the $V_{ub}$ independent  ratio $\mathcal B(B^-\to \tau^-\bar\nu_\tau)/\mathcal B(\bar B^0 \to \pi^+\ell^-\bar \nu_\ell)$ will clarify presence of new physics in $b \to u \tau \nu_\tau$ transitions. The $B\to D^* \tau \bar \nu_\tau$ mode has an unique sensitivity to the pseudoscalar density operator which does not contribute to the $B\to D \tau \nu_\tau$ decay mode, while the opposite is true for the scalar density operator.  Therefore, the precise experimental study of both $B\to D^{(*)} \tau \bar \nu_\tau$ decay modes  will be extremely useful in constraining these kinds of beyond SM physics, especially, since present branching fraction measurements of all helicity suppressed (semi)leptonic modes of B mesons are systematically above SM predictions.



\end{document}

%% file: econfmacros.tex
\def\beq{\begin{equation}}
\def\eeq#1{\label{#1}\end{equation}}
\def\eeqn{\end{equation}}

\def\beqa{\begin{eqnarray}}
\def\eeqa#1{\label{#1}\end{eqnarray}}
\def\eeqan{\end{eqnarray}}

\let\bar=\overbar

\def\Dslash{\not{\hbox{\kern-4pt $D$}}}
\def\dslash{\not{\hbox{\kern-2pt $\del$}}}

\def\msb{{\bar{\ssstyle M \kern -1pt S}}}

